\documentclass[proceedings]{JHEP3} 
\PrHEP{ hep2001}

\usepackage{epsfig,multicol}			

\def\pz{\phantom{0}}
\def\ds{{\rm D}_{\rm s}}
\def\dsf{{\rm D}_{\rm s}}
\def\dss{{\rm D}^{\star}_{\rm s}}
\def\dssf{{\rm D}^{\star}_{\rm s}}
\def\Z{{\rm Z}}
\def\fds{{\rm f}_{\rm D_{\rm s}}}
\def\mds{m_{\rm D_{\rm s}}}
\def\tds{\tau_{\rm D_{\rm s}}}
\def\dst{\ds\to\tau\nu}

\def\dsm{\ds\to\mu\nu}
\def\dsmg{\gamma\ds\to\gamma\mu^-\bar{\nu}_\mu}
\def\ccbar{\mbox{c}\overline{\mbox{c}}}
\def\csbar{\mbox{c}\overline{\mbox{s}}}

\def\ee{\mbox{e}^+\mbox{e}^-}

\def\BR{{\rm BR}}
\def\D{{\rm D}}
\def\Ds{{\rm D}_{\rm s}}
\def\X{{\rm X}}
\def\e{{\rm e}}
\def\K{{\rm K}}

\newbox\mybox
\newcommand\fverb{\setbox\mybox=\hbox\bgroup\verb}
\newcommand\fverbdo{\egroup\medskip\noindent\fbox{\unhbox\mybox}\ }
\newcommand\fverbit{\egroup\item[\fbox{\unhbox\mybox}]}
\newcommand {\downto}
        {\mbox{ \begin{picture}(14,10)
                   \put(0,10){\line(0,-1){5.0}}
                   \put(2,5){\oval(4,4)[bl]}
                   \put(1,0){\makebox(0,0)[bl]{$\rightarrow$}}
                \end{picture} }}

\title{\boldmath Experimental Status of Leptonic D$_{\rm s}$ Decays
\unboldmath}

\author{\speaker{Stefan S\"oldner-Rembold} 
	\thanks{Heisenberg Fellow of the Deutsche Forschungsgemeinschaft}\\
	CERN, CH-1211 Geneva 23, Switzerland\\
	E-mail: \email{stefan.soldner-rembold@cern.ch}}

\conference{HEP2001}

\abstract{The measurements of the leptonic branching
ratios $\BR(\ds\to\tau\nu)$ and
$\BR(\ds\to\mu\nu)$ are reviewed. The values of the $\ds$ decay
constant $\fds$ derived from the measurements are updated and
a world average is calculated taking into account the
large correlations between the measurements.}

\begin{document} 

\section{Introduction}
The branching ratio of the purely leptonic $\ds\to 
\ell^-\bar{\nu_\ell}$ decay\footnote{Charge conjugate decays are
implied throughout the paper.}
can be calculated~\cite{bib-pot} using
\begin{equation}
\label{eq-th}
\mbox{BR}(\ds\to \ell^-\bar{\nu}_\ell)=
\frac{G_{\rm F}^2 }{8\pi} \mds 
m_\ell^2\left(1-\frac{m_{\ell}^2}{\mds^2}\right)^2|V_{\rm cs}|^2\tds 
\fds^2,
\label{eq1}
\end{equation}
where $\mds$ is the mass and $\tds$ the lifetime of the $\ds$ meson, 
$\fds$ the $\ds$ decay constant and
$V_{\rm cs}$ the corresponding CKM matrix element. 
$G_{\mbox{\scriptsize F}}$ denotes the Fermi coupling constant and
$m_{\ell}$ the mass of the lepton.

Several models for the calculation of the decay constant $\fds$ 
exist: potential models predict  $\fds$ in the range 
from 129 MeV to 356 MeV~\cite{bib-pot}, QCD sum rule models 
predict $\fds=235\pm 24$~MeV~\cite{bib-sum1} and
$\fds=230\pm 24$~MeV~\cite{bib-sum2}, 
and lattice QCD calculations predict 
$\fds = 255\pm30$~MeV~\cite{bib-latqcd}.

The extraction of CKM matrix elements from 
$\mbox{B}^0-\overline{\mbox{B}}^0$ oscillation measurements
relies on these theoretical models for calculation of the decay constant for 
B mesons, $f_{{\rm B}}$, since a measurement of $f_{{\rm B}}$ from
B$^-\to\ell^-\bar{\nu}_{\ell}$ decays is currently not feasible.
It is therefore important to measure $\fds$ to
test the theoretical models used in the $f_{{\rm B}}$ calculation.

\section{LEP measurements of \boldmath$\BR(\ds\to\tau\nu)$}
Since the $\ds\to\ell\nu$ decay is helicity suppressed, 
the $\tau$ channel 
has the largest branching ratio of all leptonic channels.
Eq.~\ref{eq1} predicts the branching ratio into electrons to be negligible,
${\rm BR}(\ds\to{\rm e}{\nu})/{\rm BR}(\ds\to\tau\nu)<10^{-5}$, due
to the factor $m_{\ell}^2$,
whereas the branching ratio into muons, BR($\dsm$), is expected to be 
sizable, ${\rm BR}(\dsm)/{\rm BR}(\dst)=0.103$. 

ALEPH~\cite{bib-ALEPH} measures the signal by separating 
$\ds\to\tau\nu, \tau\to\e\nu\nu,\mu\nu\nu$ and $\ds\to\mu\nu$
decays from background using linear discriminants. 
The branching ratio measured by ALEPH is
$\BR(\Ds\to\tau\nu)=
(5.79\pm0.76~({\rm sta})\pm1.16~({\rm sys})\pm1.35~(\phi\pi))\%.$
The last error is due the uncertainty on the $\Ds$ production
rate which is dominated by the uncertainty on 
$\BR(\Ds\to\phi\pi)=(3.6\pm0.9)\%$. This uncertainty is common
to almost all measurements of leptonic $\Ds$ decays, and is
therefore treated separately.

DELPHI~\cite{bib-DELPHI}, L3~\cite{bib-l3ds} and OPAL~\cite{bib-OPAL}
have measured $\BR(\dst$) by reconstructing the decay sequence
\begin{tabbing}
\hspace{10mm} \= \hspace{33mm} \= \hspace{7mm} \= \hspace{3mm} \=
\hspace{7mm} \= \hspace{40mm}  \=  \kill 
 \>$\ee\to\Z\to\ccbar\rightarrow$\>$\dss \;\;{\rm X}$ \>                  \>           \>\> 
 \\
 \>                        \>$\downto$\> $\gamma$\> $\ds$\>\> 
 \\
 \>                        \>         \>                  \>$\downto$  \> 
$\;\tau\;\nu$ \>  \\
 \>                        \>         \>                  \>\> $\downto 
{\ell}\;\nu\;\nu\;\; (\ell=e,\mu). $\\
\end{tabbing}
\begin{flushright} \vspace{-1.3cm} (2.1) \end{flushright}
\setcounter{equation}{1}
Only $\dst$ events from $\Z\to {\rm c}\bar{\rm c}$ decays are considered,
since a measurement of BR($\dst$) in $\Z\to {\rm b}\bar{\rm b}$ events is 
systematically limited by the large uncertainty 
on the production rate of $\ds$ mesons in $\Z\to {\rm b}\bar{\rm b}$ events.

For a sample of preselected hadronic Z events with one identified
electron or muon, the kinematics are required to be consistent 
with $\dst\to\ell^-\bar{\nu}_{\ell}\nu_\tau\bar{\nu}_\tau$ decays.
In the final step of the analysis $\dss\to\gamma\ds$ decays are 
reconstructed in this $\dst$ enhanced sample by forming
the invariant mass of the photon and the $\ds$ candidate. 
This reduces the 
dependence on the Monte Carlo simulation of the background and
increases the purity of the $\ds$ sample.

The decay $\dsm$ is included in the
signal definition and the final result is corrected for this contribution.
\begin{eqnarray}
\mbox{BR}(\dst) = \frac{N_{\mbox{\scriptsize cand}}}{
2 N_{\Z}\cdot R_{\mbox{c}} \cdot f(c\to \ds) \cdot
P_V(\dssf,\dsf)
\cdot\mbox{BR}(\dss\to\gamma\ds)}
 \nonumber \\
\times\frac{1}{\mbox{BR}(\tau\to l\bar{\nu}_l\nu_\tau)\cdot\epsilon
(\dst) + \frac{{\rm BR}(\dsm)}{{\rm BR}(\dst)} 
\cdot\epsilon(\dsm)},\quad
\end{eqnarray}
$N_{\rm cand}$ is the number of background-subtracted candidates
in the signal region, $N_{\Z}$ the number of $\Z$ decays,
$R_{\rm c}=0.1729\pm0.0032$~\cite{bib-frag} 
the partial width of the $\Z$ decaying into a pair of 
charm quarks, $f(c\to \ds)=0.130\pm0.027$~\cite{bib-frag} the production rate
of $\ds$ mesons in charm jets,
$\epsilon (\dst)$ the efficiency for the signal and $\epsilon (\dsm)$ 
the efficiency for reconstructing $\dss\to\dsmg$ decays.

$P_V(\dssf,\dsf)$ is the ratio of $\csbar$ mesons produced in a 
vector state ($\dssf$) with respect to the
sum of the pseudoscalar ($\dsf$) and vector states.
For non-strange D mesons, $P_V({\rm D}^{\star},{\rm D})$ has been measured 
by ALEPH~\cite{bib-apv},
DELPHI~\cite{bib-dpv} and OPAL~\cite{bib-opv}. The averaged
value is $P_V({\rm D}^{\star},{\rm D})=0.61\pm0.03$~\cite{bib-tpv}.
To extrapolate this ratio to $\dsf$ mesons, the effect
of the decays of $L=1$ D$^{\star\star}$ resonances and quark
mass effects need to be taken into account. 
D$^{\star\star}$ resonances contribute only in the case of non-strange mesons.
This effect was estimated by OPAL
to be smaller than the experimental uncertainty~\cite{bib-opv}
and is therefore neglected.
Applying the correction factor for quark mass effects from~\cite{bib-tpv}
yields $P_V(\dssf,\dsf)=0.64\pm0.05$ where the full size of the
correction is included in the uncertainty. This value
is consistent with the ALEPH measurement of 
$P_V(\dssf,\dsf)=0.60\pm0.19$~\cite{bib-apv}.

Using these input values and $P_V(\dssf,\dsf)=0.64\pm0.05$ we obtain
the following measurements:
\begin{eqnarray}
{\rm ALEPH~}: \BR(\Ds\to\tau\nu)=
(5.79\pm0.76~({\rm sta})\pm1.16~({\rm sys})\pm1.35~(\phi\pi))\%\pz\pz\pz\pz\pz\pz\pz\pz\pz\pz\pz
\\ \nonumber
{\rm DELPHI}: \BR(\Ds\to\tau\nu)=
(6.91\pm3.45~({\rm sta})\pm1.72~({\rm sys})\pm1.43~(\phi\pi)\pm0.55~(P_V))\%
\\ \nonumber
{\rm L3~~~~}: \BR(\Ds\to\tau\nu)=
(6.34\pm2.44~({\rm sta})\pm1.38~({\rm sys})\pm1.32~(\phi\pi)\pm0.51~(P_V))\%
\\ \nonumber
{\rm OPAL~~}: \BR(\Ds\to\tau\nu)=
(6.25\pm1.91~({\rm sta})\pm1.12~({\rm sys})\pm1.30~(\phi\pi)\pm0.50~(P_V))\%
\end{eqnarray}
The average is
$\BR(\Ds\to\tau\nu)=
(6.05\pm1.04\pm1.34~(\phi\pi)\pm0.22~(P_V))\%.$

\section{Measurements of \boldmath$\BR(\ds\to\mu\nu)$\unboldmath}
Before LEP several experiments have measured the branching ratio
of the decay D$_{s}\to\mu\nu$ to derive $\fds$. 
These measurements also depend
on external input which is partially correlated. In the following
I will therefore shortly review the measurements and give an updated
result wherever external inputs have changed.
\begin{itemize}
\item 
The WA75 experiment~\cite{bib-wa75} has used
$350$~GeV $\pi$ nucleon interactions with an emulsion target
to measure the ratio
\begin{equation}
r \frac{\BR(\Ds\to\mu\nu)}{\BR(\D^0\to\mu\nu\X)}=
(1.25^{+0.55}_{-0.44}~(\rm sta)^{+0.24}_{-0.20}~(sys))\cdot 10^{-2},
\end{equation} 
where $r$ is ratio of the production cross-section for D$_{\rm s}$
and D$^0$ mesons in $\pi$n scatttering
The ratio $r$ can be derived
from a BEATRICE measurement of these cross-section
in the forward direction $(x_{\rm F}>0)$ to be 
$r=0.166\pm0.026\pm0.041~(\phi\pi)$~\cite{bib-bea2}~\footnote{The
correlation introduced by using the same BEATRICE 
$\ds\to\K^+\K^-\pi$ data to normalise the BEATRICE and the
WA75 measurement is found to have a negligible effect on the combined
result}. 
With $\BR(\D^0\to\mu\nu\X)=0.066\pm0.008$~\cite{bib-pdg} we obtain 
\begin{equation}
\BR(\Ds\to\mu\nu)
=(0.50^{+0.22}_{-0.18}~({\rm sta})^{+0.14}_{-0.13}~({\rm sys})\pm 
0.12({\phi\pi}))\%.
\end{equation}
\item 
Using interactions of $350$~GeV $\pi^-$ on copper and tungsten
targets, the BEATRICE experiment has measured the ratio~\cite{bib-bea1}  
\begin{equation}
\frac{\BR(\Ds\to\mu\nu)}{\BR(\Ds\to\phi(\to\K^+\K^-)\pi)}=
0.47\pm0.13~({\rm sta})\pm0.04~({\rm sys})\pm0.06~(\phi\pi) 
\end{equation} 
which yields
$\BR(\Ds\to\mu\nu)=
0.83\pm0.23~({\rm sta})\pm0.06~({\rm sys})\pm0.18~(\phi\pi).$
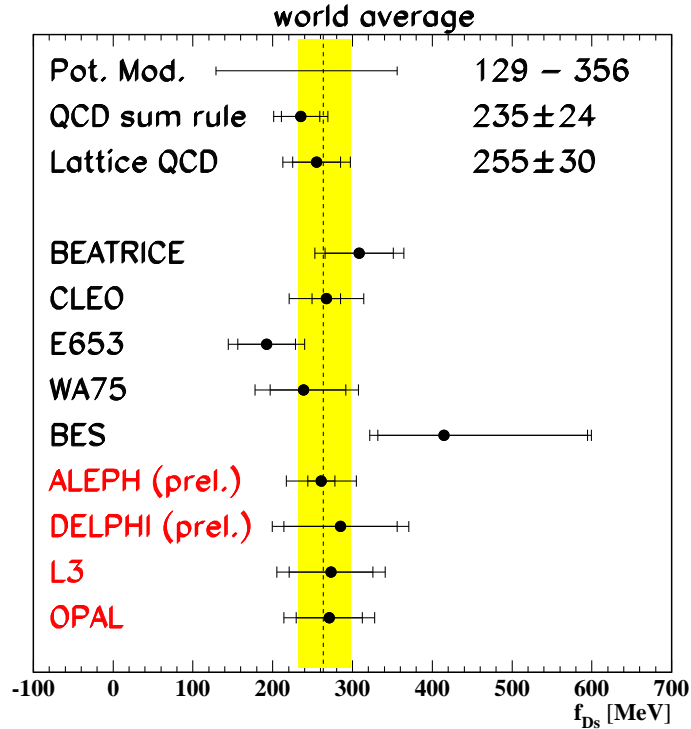
\EPSFIGURE[htb]{fds_comp.epsi,width=9.0cm}{Comparison of
the $\fds$ measurements by the individual experiments with
the theoretical predictions. The inner error bar is
the statistical uncertainty and the outer error bar
the total uncertainty.
The dashed line is the world average calculated in this note
and the yellow band the total uncertainty.
}
\item
The E653 experiment~\cite{bib-e653} has measured the ratio
\begin{equation}
\frac{\BR(\Ds\to\mu\nu)}{\BR(\Ds\to\phi\mu\nu)}=
0.16 \pm 0.06~({\rm sta}) \pm 0.03~({\rm sys}) 
\end{equation}
in $600$~GeV $\pi$ nucleon interactions on an emulsion.
Using $\BR(\Ds\to\phi\mu\nu)=0.020\pm0.005$~\cite{bib-pdg}
this yields 
$\BR(\Ds\to\mu\nu)
=(0.32\pm0.12~({\rm sta})\pm0.07~({\rm sys})\pm0.08~(\phi\pi))\%$.
\item 
The BES experiment~\cite{bib-bes} has measured 
$\BR(\Ds\to\mu\nu)=
(1.5^{+1.3}_{-0.6}~({\rm sta})^{+0.3}_{-0.2}~({\rm sys}))\%$.
in the process $\e^+\e^-\to\Ds\Ds$ by tagging leptonic $\Ds$ decays
recoiling to a hadronic $\Ds$ decay. The uncertainty is mainly statistical
and no correlation needs to be taken into account.
\item The most recent CLEO measurement~\cite{bib-cleo1} of the ratio
\begin{equation}
\frac{\BR(\Ds\to\mu\nu)}{\BR(\Ds\to\phi\pi)}=
0.173 \pm 0.023~({\rm sta})  \pm 0.035~({\rm sys})  
\end{equation} 
is based on $\ee\to\ccbar$ events measured at energies
close to the $\Upsilon(4S)$ resonance.
The branching ratio is derived to be
$\BR(\Ds\to\mu\nu)
=(0.62\pm0.08~({\rm sta})\pm0.13~({\rm sys})\pm0.16~(\phi\pi))\%$.
\end{itemize}
\TABLE[ht]{
\begin{tabular}{|l|l|c|}
\hline
Experiment     & $\pz\pz\pz\fds$~(MeV) & $\fds$~(MeV) \\ \hline
ALEPH (prel.)  & $261 \pm 17 \pm 26 \pm 30  $   
               & $285 \pm 20 \pm 40 $  \\
DELPHI (prel.) & $285 \pm 71 \pm 35 \pm 30 \pm 11$   
               & $330 \pm 82 \pm 50 $  \\ 
L3             & $273 \pm 52 \pm 30 \pm 29 \pm 11$
               & $309 \pm 58 \pm 50 $  \\ 
OPAL           & $271 \pm 41 \pm 24 \pm 28 \pm 11$
               & $286 \pm 44 \pm 41 $  \\
Beatrice       & $309 \pm 43 \pm 11 \pm 33 $   & $323    \pm 44  \pm 36$  \\
CLEO           & $267 \pm 18 \pm 27 \pm 33 $   & $280    \pm 19  \pm 44$  \\
E653           & $192 \pm 36 \pm 20 \pm 24 $   & $194    \pm 35  \pm 24$  \\
WA75           & $239^{\pz\pz+53~\pz+33}_{\pz\pz-42~\pz-30}\pm 30$   
                                           & $232    \pm 45  \pm 52$  \\
BES            & $418^{\pz\pz+180~+41}_{\pz\pz-83~\pz-28}$ &
                               $430^{~+150}_{~-130}\pm 40$ 
\\
\hline
\end{tabular}
\label{tab1}
\caption{Decay constants $\fds$ measured by the experiments.
The first value shown is calculated with the numbers given
in the paper. The uncertainties are the statistical uncertainties,
the uncorrelated systematic uncertainties, the uncertainties due
to $\BR(\ds\to\phi\pi)$, and the uncertainties due to $P_V$.
The second value is the original value published by the experiment with
the statistical and the sum of all systematic uncertainties.
The relative uncertainties of the BES measurement are different
because in the original analysis $\fds$ has been extracted directly
from the data~\protect\cite{bib-bes}.}
}
Averaging these results yields
$\BR(\Ds\to\mu\nu)=(0.53\pm0.09\pm0.12~(\phi\pi))\%$ taking into
account the correlation due to the uncertainty on the
branching ratio $\BR(\Ds\to\phi\pi)$. 
The first error is due to the
uncorrelated statistical and systematic uncertainties. 
The purely statistical contribution to the uncertainty is 10~MeV.
The uncorrelated
uncertainties yield $\chi^2/{\rm ndf}=6.5/4$. 
The result is in
good agreement with the ALEPH measurement~\cite{bib-ALEPH}
$\BR(\Ds\to\mu\nu)=(0.68\pm0.16\pm0.13~(\phi\pi))\%$ which
is not used in the averages due to its large correlations
with the ALEPH $\ds\to\tau\nu$ measurement.
\section{Decay constant \boldmath $\fds$ \unboldmath}
The decay constant $\fds$ is calculated using (\ref{eq1}) with
$G_{\rm F}=(1.16639 \pm 0.00001)\times 10^{-5}$~GeV$^{-2}$,
$|V_{\rm cs}|=0.9891 \pm 0.016$,
$\mds=1.9686 \pm 0.0006$ GeV,
$\tds=(0.496 \pm 0.01)\times 10^{-12}\, {\rm s}$,
$m_\tau=1.77703 \pm 0.00030$~GeV~\cite{bib-pdg}.  
Most uncertainties are negligible,
only the uncertainties on $|V_{\rm cs}|$ and $\tds$ contribute slightly
to the final uncertainty.

In table~\ref{tab1} the values of $\fds$ are compared to the original
values published by the experiments. 
The average of all measurements yields
\begin{equation}
\fds=264\pm 15\pm 33~(\phi\pi) \pm 2~(P_V) \pm 4~(V_{\rm cs},\tds)~\mbox{MeV},
\end{equation}
where the first uncertainty is due to the sum of the statistical and
the uncorrelated systematic uncertainties of the measurements, and 
the other uncertainties are due to the various correlated uncertainties.
Using only the uncorrelated uncertainties yields 
$\chi^2/{\rm ndf}=7.4/8$.

In figure 1 the $\fds$ measurements of the experiments
are shown together with the average calculated in this note.
They are also compared to the theoretical predictions. Within
the uncertainties they are consistent with the data. The precision
of the measurement can only be increased by reducing
the uncertainty on $\BR(\ds\to\phi\pi)$ or by using
measurement methods which do not depend on $\BR(\ds\to\phi\pi)$.

\acknowledgments
I would like to thank Christian Schmitt (Wuppertal) for his
help in preparing this note.

\end{document}